\documentclass[pra,showpacs,twocolumn,floatfix]{revtex4}%
\usepackage{amsmath}
\usepackage{graphicx}
\usepackage{amsfonts}
\usepackage{subfigure}
\usepackage{graphicx}
\usepackage{float}

\usepackage{amssymb}%

\newcommand{\ket}[1]{\ensuremath{\left\vert #1 \right\rangle}}

\begin{document}
\title{Light storage in a magnetically-dressed optical lattice}
\date{\today }
\author{Y. O. Dudin, R. Zhao, T.A.B. Kennedy, and A. Kuzmich}
\affiliation{School of Physics, Georgia Institute of
Technology, Atlanta, Georgia 30332-0430\\}  

\begin{abstract}
Differential Stark shift compensation for ground state $^{87}$Rb atoms trapped in an elliptically polarized optical lattice and ``magic" magnetic field, was recently proposed and demonstrated experimentally by N. Lundblad {\it et al.}, arXiv:0912.1528 and analyzed theoretically by A. Derevianko, arXiv:0912.3233.  Here we demonstrate for the first time enhanced hyperfine coherence times using the magic field technique. We observe coherent light storage with a 0.32 s lifetime in an atomic Rb gas confined in a one dimensional optical lattice and magnetic field.
\end{abstract}

\maketitle

Achieving long lifetimes for atomic coherence is of major importance for precision measurements and quantum computing and communication
tasks. All of these require elimination of decoherence arising from inhomogeneous external fields and trapping potentials. In atomic fountains long coherence times have been observed using clock transitions, which are first order insensitive
to magnetic fields \cite{ashby}. However for quantum information processing it is important to control the positions of atoms. A common strategy involves the use of an optical lattice or an array of micro-traps \cite{jessen,yavuz,kruse}. Ground level coherences, for atoms at different positions in the lattice, will accumulate different phases due to the differential ac-Stark shift caused by the lattice field. This results in qubit dephasing, on millisecond timescales, for protocols in which cold, but non-degenerate, atoms are distributed across the lattice \cite{kuhr}. Spin-wave (collective) qubits have also been stored in a 1-D optical lattice, with lifetime of 7 ms limited by ac-Stark decoherence \cite{zhao,dudin}. The differential light shifts for degenerate atomic gases can be decreased by employing lower trap depths. For instance, 0.24 second lifetime for storage of classical light pulses has been reported in \cite{schnorr}, albeit with a low (0.3\%) storage efficiency due to the small atomic sample. Low numbers of atoms also make it difficult, if not impossible, to carry out protocols, such as Duan-Lukin-Cirac-Zoller, central to the quantum repeater implementations \cite{duan}.

For optical transitions in alkaline-earth atoms there exist special wavelengths of the lattice light at which the ac-Stark shifts for ground and excited states are identical. This is used in realizing high accuracy optical clocks \cite{katori}. In contrast, complete elimination of the ac Stark shift for the microwave ground level coherences of alkali atoms in optical lattices or micro-trap arrays did not appear to be feasible.

Recently however, laser-induced compensation was proposed and employed to demonstrate light storage and retrieval in a 1-D optical lattice for lifetime of $\sim 0.2$ seconds \cite{radnaev}. An ac differential Stark shift-free lattice based on elliptically polarized light fields has been proposed \cite{flambaum,cho}. However, the scheme applies to hyperfine coherences which are first-order sensitive to magnetic fields, and therefore requires an extremely high degree of magnetic field control. More recently, Lundblad et al. have proposed, and demonstrated spectroscopically, that by mixing the two ground hyperfine levels with a dc magnetic field, it is possible to eliminate ac-Stark shift on the clock transition in an elliptically polarized lattice at a particular (magic) magnetic field value \cite{lundblad}. While the clock coherence becomes first-order sensitive to magnetic field, it is weakly so, and this is promising for greatly increased coherence times.

Here we report the first measurements of increased clock coherence times using this technique. We achieve coherent light storage and retrieval in the atomic sample of $\sim 10^7$ rubidium atoms confined in a 1-D optical lattice at a temperature of $\sim 10$ $\mu$K, with lifetime of 0.32 seconds, bettering previous results achieved via laser compensation \cite{radnaev} and in a shallow-lattice Mott insulator \cite{schnorr} by about 30\%. The storage and retrieval efficiency of $3.4\%$ (at a storage time of 1 ms) is comparable to that of Ref.\cite{radnaev}, but is more than an order of magnitude higher than in the Mott insulator work \cite{schnorr}. We also observe long storage times, albeit with ten times lower efficiency, employing the $|F=2,m=\pm 1\rangle \leftrightarrow |F=1,-m\rangle$ hyperfine coherences (we will employ the shorthand $\pm1/\mp1$). By comparing results for the three distinct coherences, we establish residual magnetic sensitivity of the approach as a major limiting factor on the observed storage lifetimes, and identify differential vector and rank-2 tensor ac Stark contributions to the measured magnetic field values.

We load $^{87}$Rb atoms into a magneto-optical trap (MOT) from a background vapor. After compression and precooling the sample is transferred to a one dimensional optical lattice resulting in an approximately cigar shaped cloud of $\sim 10^7$ atoms with $1/e^2$ waists of 130 $\mu$m and 840 $\mu$m, respectively. The lattice is formed by interfering two circularly polarized 1063.8 nm (in vacuum) beams intersecting at an angle of 9.6$^{\circ}$ in the horizontal plane. The beams have waists of $\sim 200$ $\mu$m and their total power is varied between 8 W and 14.6 W.

The probe and control fields are tuned to corresponding D$_1$-resonance lines $\ket{a}\leftrightarrow \ket{c}$ and $\ket{b}\leftrightarrow \ket{c}$, respectively, where $\{\ket{a,b}\}$ correspond to $\{5S_{1/2}, F=2, 1\}$ and $\{\ket{c}\}$ represents $\{5P_{1/2}, F=2\}$.
The probe and control beams co-propagate at a small angle of 0.9$^\circ$, and are positioned in the horizontal plane symmetrically with respect to the lattice and magnetic field direction, cf Refs.\cite{zhao,dudin}. The beam waists are 110 and 270 $\mu$m, respectively. The non-zero angle results in a phase grating imprinted on the atoms whose spatial profile is preserved from motional dephasing by the lattice.

Three different polarization configurations of the probe and control fields are employed, each corresponding to excitation of a different long-lived hyperfine coherence: lin$\perp$lin for clock coherence,
$\sigma^{\pm}/\sigma^{\mp}$ for $\mp 1/\pm 1$ coherences. After loading, a bias magnetic field is applied along the major axis of the trap and atoms are either prepared in the 5$^2S_{1/2}, F = 1, m = 0$ state by means of optical pumping when clock coherence is addressed, or left unpolarized when $\pm1/\mp 1$ coherences are used.

The probe pulse is adiabatically converted into the respective spin wave that is first-order insensitive to ambient magnetic fields. The probe field has a full width at half maximum (fwhm) of 60 ns and peak power of 10 nW. The control field, also used for retrieval, has fwhm of 330 ns and peak power 90 $\mu$W. After a given storage period the control field retrieves a signal on the probe transition. The latter is coupled to a single mode fiber and directed onto an avalanche photo-detector, the voltage output of which is digitized and recorded. We measure the retrieval efficiency, defined as the ratio of the energies of the retrieved and incident probe pulses after a given storage time.
Magnetic field strength is calibrated by measurements of the Larmor precession frequencies of the stored spin-waves in an unpolarized sample of atoms (no optical pumping) \cite{jenkins,matsukevich2}. The lattice is linearly polarized during the calibration to exclude the contribution of vector lattice light shift to the Larmor frequency. Further details of the experimental setup are given in Refs. \cite{zhao,dudin}.

\begin{figure}
  \centering
  \includegraphics[width=3.0in]{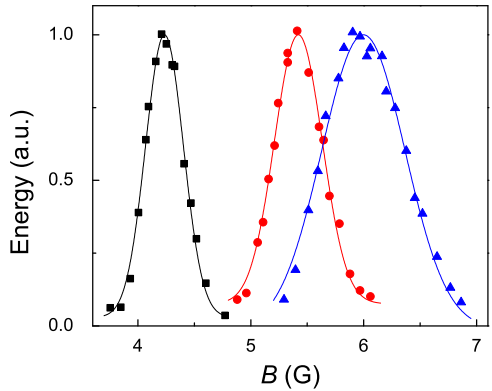}
  \vspace{-0.2cm}
  \caption{(Color
  online) Retrieved pulse energy as a function of the magnetic field, for 64 $\mu$K deep lattice. Squares, for lin$\perp$lin polarizations and $m=0$ optically pumped sample, circles and triangles for the $\sigma^{\pm}/\sigma^{\mp}$ polarizations, unpolarized sample. Light is stored and retrieved after 0.5, 0.3 and 0.2 s, respectively. Solid curves are Gaussian fits, see text.}
  \label{fig.Bmax}
\end{figure}

\begin{figure}[h]
  \centering
  \includegraphics[width=3.0in]{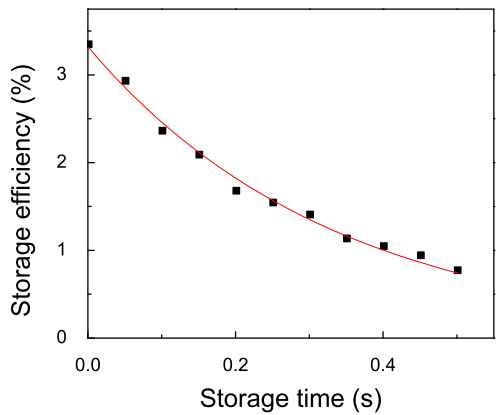}
  \vspace{-0.2cm}
  \caption{(Color
  online) Retrieval efficiency vs storage time in a 48 $\mu$K deep lattice at $B=4.2$ G. An exponential fit provides lifetime of 0.32(1) second. See text for details.}
  \label{fig.storage}
\end{figure}

\begin{figure}
  \centering
  \includegraphics[width=3.0in]{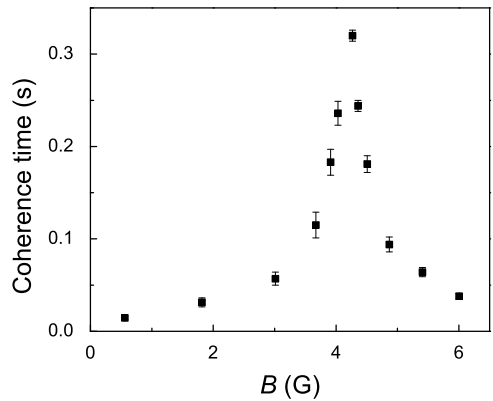}
  \vspace{-0.2cm}
  \caption{(Color
  online) 1/e lifetime determined via the storage and retrieval protocol as a function of the magnetic field in a 64 $\mu$K deep lattice.}
  \label{fig.cohtime}
\end{figure}

The ac-Stark shift can be described by an operator $U = \sum_{L=0,1,2} U^{[L]}$, with scalar, vector and (rank-2) tensor parts each proportional to the light intensity.
In a circularly polarized lattice with the magnetic field directed along the propagation/quantization direction, both the Zeeman interaction $H^Z$ and $U$ couple only hyperfine states with the same $m$. Since the ground level hyperfine splitting is large
compared to $H^Z$ and $U$, we may compute the level shifts using perturbation theory, giving, for $m=\pm 1$, the differential shift $\hbar \delta_m$ associated with the energy difference between the states $|F=2,m=\pm 1\rangle$ and  $|F=1,-m\rangle$
\begin{widetext}
\begin{eqnarray*}
\hbar \delta_m &&\simeq
 \langle 2 m | H^Z +  U^{[0]}+U^{[1]} +U^{[2]}|2 m \rangle -\langle 1 -m | H^Z + U^{[0]}+U^{[1]} +U^{[2]} |1 -m \rangle +\\ &&\frac{ \langle 2 m | H^{Z}+U^{[1]} |1 m \rangle \langle 1 m | H^{Z}+U^{[1]} |2 m\rangle + \langle 1 -m | H^{Z}+U^{[1]} |2 -m \rangle
\langle 2 -m | H^{Z}+U^{[1]} |1 -m\rangle }{\hbar \Delta}
\end{eqnarray*}
\end{widetext}
where  $\hbar \Delta$ is the ground level hyperfine splitting. We have ignored the small tensor contribution in the second-order term, and used the fact that the scalar shift does not couple the $F=1$ and $2$ states by symmetry.

Suppressing the term linear in light intensity largely removes the dephasing due to the optical lattice. This occurs at the field value
\begin{widetext}
\begin{eqnarray}
\mu B_0^{(\pm)} && =\frac{4}{3} \frac{\hbar\Delta}{4\alpha_{12}^{[1]}} \big[\sqrt{\frac{5}{3}}(\alpha_2^{[0]}-\alpha_1^{[0]})\pm
\frac{\sqrt{5}}{2}(\alpha_1^{[1]}+{\frac{1}{\sqrt{3}}}\alpha_2^{[1]})-
\frac{1}{2}\sqrt{\frac{5}{21}}(\alpha_2^{[2]}+\sqrt{\frac{7}{5}}\alpha_1^{[2]})\big].
\end{eqnarray}
\end{widetext}

A similar calculation for the clock coherence results in the corresponding field value

%\begin{widetext}
\begin{equation}
\mu B_0^{(0)}  =\frac{\hbar\Delta}{4\alpha_{12}^{[1]}} \big[\sqrt{\frac{5}{3}}(\alpha_2^{[0]}-\alpha_1^{[0]})
-\sqrt{\frac{5}{21}}(\alpha_2^{[2]}-\sqrt{\frac{7}{5}}\alpha_1^{[2]})\big].
\end{equation}
%\end{widetext}

The diagonal tensor polarizabilities $\alpha^{[p]}_{F}$ for $F=1,2$ correspond to $\alpha_{p}$ of Ref.\cite{manakov} in which the ground-state energy is that of the hyperfine level of spin $F$, whereas $\alpha_{12}^{[p]}$ is an off-diagonal generalization associated with the Stark coupling between the hyperfine ground levels and is similar to that discussed in Ref.\cite{derevianko}. In Eqs. 1 and 2 the scalar contribution dominates the measured magnetic field values. However, suitable combinations of the latter enable us to extract the vector and rank-2 tensor contributions, as we will discuss later.

In order to determine $B_0$, we maximize the retrieved pulse energy while tuning the magnetic field strength, as shown in Fig.\ref{fig.Bmax}. The data are fitted with the function $\exp (-\gamma (B-B_0)^2)$,
$\gamma$ being an adjustable parameter, resulting in $B_0^{(0)}$=4.24(1) G, $B_0^{(+)}$=5.42(1) G, and $B_0^{(-)}$=5.99(2) G, respectively, where the 0.01 G uncertainty is the combined statistical and field calibration error.

Accounting for the measured degree of circular polarization $A=0.991(2)$ of the lattice beams \cite{derevianko,A}, the value of the magnetic field in an ideal circularly-polarized lattice collinear with the magnetic field would be $B_0^{(0)}$=$4.20(1)$ G, while Ref.\cite{derevianko} predicts the value of $B_0^{(0)}$ = 4.38 G, for our 1063.8 nm lattice. The disagreement deserves further investigation.

The similarly corrected values of $B_0^{(+)}$ and $B_0^{(-)}$, for perfect circular polarizations, would be 5.37(1) G and 5.93(2) G, respectively. Their (normalized) difference $2[B_0^{(-)}-B_0^{(+)}]/[B_0^{(-)} + B_0^{(+)}]=0.099(3)$ is proportional to the differential vector shift, see Eq.(1) and is consistent with an estimated value of 0.11 when only the 5S and 5P levels are included in the calculation of the ac Stark shifts. A non-zero value of the quantity $[(3/8)(B_0^{(-)} + B_0^{(+)})-B_0^{(0)}]/B_0^{(0)} = 0.008(3)$ reveals ac Stark shifts associated with the rank-2 tensor, and is consistent with the value of 0.009 estimated by considering only the 5S and 5P levels.

\begin{figure}
  \centering
  \includegraphics[width=3.0in]{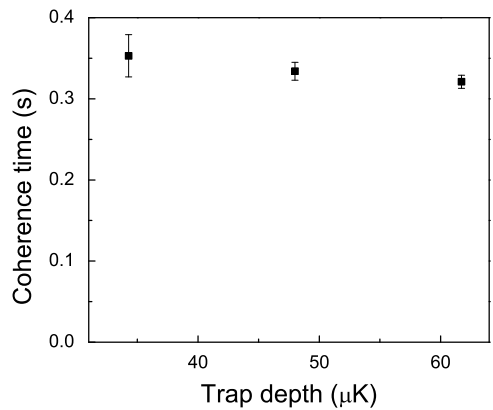}
  \vspace{-0.2cm}
  \caption{(Color
  online) The 1/e lifetime of light storage and retrieval protocol vs trap depth. Memory lifetime is no longer limited by Stark shifts but magnetic decoherence of the clock states and atomic losses.}
  \label{fig.trapdepth}
\end{figure}

\begin{figure}
  \centering
  \includegraphics[width=3.0in]{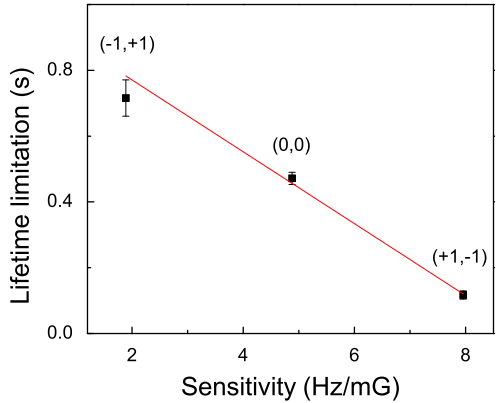}
  \vspace{-0.2cm}
  \caption{(Color online) $T \tau_{c,\pm}/(T -\tau_{c,\pm})$ as the function of the effective magnetic field moment $\mu^{\prime}\equiv dE/dB$, for the three long-lived coherences, each at its respective magic field value $B_0^{\pm,c}$, in a 64 $\mu$K deep trap.}
  \label{fig.sensitivity}
\end{figure}

 The decay of the retrieved signal is well described by an exponential function of storage time, Fig. \ref{fig.storage}. The 1/e storage lifetimes $\tau$ versus applied magnetic field data are shown in Fig. \ref{fig.cohtime}. No significant change in lifetime is observed when the lattice trap depth is varied, as shown in Fig.\ref{fig.trapdepth}. The other lifetimes were $\tau_+ =0.43(2)$ s and $\tau_- =0.10(1)$ s. These can be explained by a combination of the 1 s lifetime of the atoms in the lattice and of the concomitant effective magnetic moments $\mu^{\prime}$, of the coherences, $\mu^{\prime}\equiv dE/dB$. The lifetime of a coherence $\tau$ in a gradient of the ambient magnetic field $B^{\prime}$ is expected to be inversely proportional to $\mu^{\prime}$:  $\tau^{-1} = 2\pi \mu^{\prime} B^{\prime} l$, where $l$ is the length of the atomic sample, e.g., see Methods section of Ref.\cite{zhao}. After excluding the atomic loss contribution to the measured storage lifetime via $1/\tau_{c,\pm}=1/\tau^m_{c,\pm}+1/T$, the residual lifetime $\tau^m_{c,\pm} \equiv T \tau_{c,\pm}/(T -\tau_{c,\pm})$ is displayed in Fig. \ref{fig.sensitivity}, and is in reasonable agreement with the expected linear dependence.

In conclusion, we have achieved a combination of high efficiency and exceptionally long storage time for coherent light pulses in an ac Stark compensated optical lattice. The measured value of the magic magnetic field $B_0=4.20(1)$ G for the clock coherence disagrees with the theoretically predicted value of 4.38 G \cite{derevianko}.  We measure long light storage lifetimes at field values 5.37(1) G and 5.93(2) G for the $\pm1/\mp1$ coherences, revealing small differential vector and rank-2 tensor light shifts. The measured lifetimes of light storage for the three coherences, of 0.32 s, 0.43 s, and 0.1 s, respectively, are consistent with dephasing in ambient magnetic fields. We expect similar lifetimes for single photon storage given that a 0.1 s single photon storage has recently been achieved with a different light shift compensation scheme \cite{radnaev}.

We thank D.N. Matsukevich for his contributions, and A. Derevianko for discussions. This work was supported by the Air Force Office of Scientific Research and the National Science Foundation.


\begin{thebibliography}{99}
\bibitem{A} Parameter $A\equiv\sqrt{1-\epsilon^2}$, where ellipticity $\epsilon$ is defined as in Ref. \cite{kuppens}.
\end{thebibliography}
\end{document}